\documentclass[preprint]{aastex}

\usepackage{emulateapj5}
\usepackage{apjfonts}

\shortauthors{METCALFE WINGET \& CHARBONNEAU}
\shorttitle{CONSTRAINTS ON $^{12}$C$(\alpha,\gamma)^{16}$O FROM
WHITE DWARF SEISMOLOGY}

\submitted{Accepted for publication in the Astrophysical Journal}

\begin{document}

\title{Preliminary Constraints on $^{12}$C$(\alpha,\gamma)^{16}$O from
White Dwarf Seismology}

\author{T.S. Metcalfe\altaffilmark{1,2} D.E. Winget\altaffilmark{1}
and P. Charbonneau\altaffilmark{3}}

\altaffiltext{1}{Department of Astronomy, Mail Code C1400, University of
Texas, Austin, TX 78712}

\altaffiltext{2}{Visiting Scientist, High Altitude Observatory, National
Center for Atmospheric Research}

\altaffiltext{3}{High Altitude Observatory, National Center for
Atmospheric Research, P.O. Box 3000, Boulder, CO 80307-3000}

\begin{abstract}

For many years, astronomers have promised that the study of pulsating
white dwarfs would ultimately lead to useful information about the physics
of matter under extreme conditions of temperature and pressure. In this
paper we finally make good on that promise. Using observational data from
the Whole Earth Telescope and a new analysis method employing a genetic
algorithm, we empirically determine that the central oxygen abundance in
the helium-atmosphere variable white dwarf GD~358 is $84 \pm 3$ percent.
We use this value to place preliminary constraints on the $^{12}{\rm
C}(\alpha ,\gamma )^{16}{\rm O}$ nuclear reaction cross-section. More
precise constraints will be possible with additional detailed simulations. 
We also show that the pulsation modes of our best-fit model probe down to
the inner few percent of the stellar mass. We demonstrate the feasibility
of reconstructing the internal chemical profiles of white dwarfs from
asteroseismological data, and find an oxygen profile for GD~358 that is
qualitatively similar to recent theoretical calculations. 

\end{abstract}


\section{INTRODUCTION}

White dwarf stars are the end-points of stellar evolution for the majority
of all stars, and their composition and structure can tell us about their
prior history. Compared to main-sequence stars, white dwarfs are
relatively simple objects to model. Once they pass through the hot
planetary nebula nucleus phase, their evolution is a slow cooling process
largely uncomplicated by nuclear burning or other poorly understood
effects. Nature has been kind by providing several varieties of otherwise
normal white dwarf stars that undergo continuous and remarkably stable
pulsations. We can determine the internal structure of these pulsating
white dwarfs by observing their variations in brightness over time and
then matching these observations with a computer model which behaves the
same way. 

In the past decade, the observational aspects of white dwarf seismology
have been addressed by the development of the Whole Earth Telescope
\citep{nat90}. This instrument is now mature, and has provided a wealth of
seismological data on the different varieties of pulsating white dwarf
stars. In an effort to bring the analytical treatment of these data to a
comparable level of sophistication, we have recently developed an
objective global optimization method for white dwarf pulsation models
based on a genetic algorithm [see \S 2 of \cite{mnw00} and references
therein for a detailed discussion of genetic algorithms]. \cite{mnw00}
applied this new approach to observations of the helium-atmosphere
variable white dwarf GD~358, allowing the stellar mass ($M_{*}$), the
effective temperature ($T_{\rm eff}$), and the mass of the surface helium
layer ($M_{\rm He}$) to be free parameters. They repeated the procedure
for several combinations of core composition and internal chemical
profiles.  Among other things, this initial study demonstrated that the
fits were significantly improved by searching globally with different core
compositions. The obvious next step is to extend this method to treat the
central oxygen abundance as a free parameter. This has the exciting
potential to place meaningful constraints on the $^{12}$C$(\alpha ,\gamma
)^{16}$O nuclear reaction cross-section. 

The value of the $^{12}$C$(\alpha ,\gamma )^{16}$O cross-section at
stellar energies is presently the result of an extrapolation across eight
orders of magnitude from laboratory data \citep{fow86}. During the final
stages of a red giant star, this reaction competes with the
triple-$\alpha$ process for the available $\alpha$-particles. As a
consequence, the final ratio of carbon to oxygen in the core of a white
dwarf is a measure of the relative rates of the two reactions
\citep{buc96}. The triple-$\alpha$ reaction is relatively well determined
at the relevant energies, so the constraint that comes from this
measurement is much more precise than other methods.

\section{FORWARD\ MODELING \label{FWDSEC}}

The initial 3-parameter fits to GD~358 \citep[hereafter MNW]{mnw00} made
it clear that both the central oxygen abundance and the shape of the
internal chemical profile should be treated as free parameters. We
modified our code to allow any central oxygen mass fraction ($X{\rm _O}$)
between 0.00 and 1.00 with a resolution of 1 percent. To explore different
chemical profiles we fixed $X{\rm _O}$ to its central value out to a
fractional mass parameter ($q$) which varied between 0.10 and 0.85 with a
resolution of 0.75 percent. From this point, we forced $X{\rm _O}$ to
decrease linearly in mass to zero oxygen at the 95 percent mass point.

This parameterization is a generalized form of the ``steep'' and
``shallow'' profiles used in \cite{mnw00}. We originally used these
profiles so that our results could be easily compared to earlier work by
\cite{woo90} and \cite{bww93}. The latter authors define both profiles in
their Figure 1. The ``shallow'' profile corresponds approximately to
$q=0.5$, and ``steep'' corresponds roughly to $q=0.8$. However, in our
generalized parameterization we have moved the point where the oxygen
abundance goes to zero from a fractional mass of 0.9 out to a fractional
mass of 0.95. This coincides with the boundary in our models between the
self-consistent core and the envelope, where we describe the He/C
transition using a diffusion equilibrium profile from the method of
\cite{af80} with diffusion exponents of $\pm3$. We do not presently
include oxygen in the envelopes, so the 

\begin{table*}
\begin{center}
\tablenum{1}\label{tab1}
\centerline{\sc Table 1}
\centerline{\sc Convergence of the method on synthetic data.}
\vskip 5pt
\begin{tabular}{cccccc}
\hline\hline
Iteration & $T_{\rm eff}$ & $M_*/M_{\odot}$ & $\log(M_{\rm He}/M_*)$ & $X_{\rm O}$ & $q$ \\ 
\hline
1$^a$ & 23,600 & 0.600 & $-$5.76 & 0.52 & 0.55 \\
1$^b$ & 22,200 & 0.660 & $-$2.79 & 0.99 & 0.55 \\
2$^a$ & 22,200 & 0.660 & $-$2.79 & 0.88 & 0.51 \\
2$^b$ & 22,600 & 0.650 & $-$2.74 & 0.85 & 0.51 \\
3$^a$ & 22,600 & 0.650 & $-$2.74 & 0.80 & 0.50 \\
3$^b$ & 22,600 & 0.650 & $-$2.74 & 0.80 & 0.50 \\
\hline\hline
\end{tabular}
\end{center}
\vskip -0.1in
\hskip 2in{$^a$ Value of $M_*/M_{\odot}$ fixed during this iteration.}

\hskip 2in{$^b$ Value of $q$ fixed during this iteration.}
\end{table*}

\noindent mass fraction of oxygen must drop to zero by this point. 

We calculated the magnitude of deviations from the mean period spacing for
models using our profiles compared to those due to smooth profiles from
recent theoretical calculations by \cite{sal97}. The smooth theoretical
profiles caused significantly larger deviations, so we conclude that the
abrupt changes in the oxygen abundance resulting from our parameterization 
do not have an unusually large effect on the period spacing (see Appendix
for details). Although the 
actual chemical profiles will almost certainly differ from the profiles 
resulting from our simple parameterization, we should still be able to 
probe the gross features of the interior structure by matching one or 
another linear region of the presumably more complicated physical profiles. 

We used the same ranges and resolution for $M_{*}$, $T_{\rm eff}$, and
$M_{\rm He}$ as in MNW, so the search space for this 5-parameter problem
is 10,000 times larger than for the 3-parameter case. Initially we tried
to vary all 5 parameters simultaneously, but this proved to be impractical
because of the parameter-degeneracy between $M_{*}$ and $q$. The pulsation
properties of the models depend on the {\it radial} location in the
chemical profile where the oxygen mass fraction begins to change. If the
genetic algorithm (GA) finds a combination of $M_{*}$ and $q$ that yields
a reasonably good fit to the data, most changes to either one of them by
itself will not improve the fit. As a consequence, simultaneous changes to
both parameters are required to find a better fit, and since this is not
very probable the GA must run for a very long time.  Tests on synthetic
data for the full 5-parameter problem yielded only a 10 percent
probability of finding the input model even when we ran for 2000
generations---ten times longer than for the 3-parameter case. By contrast,
when we used a fixed value of $q$ and repeated the test with only 4 free
parameters, the GA found the input model in only 400 generations for 8 out
of 10 runs.  Even better, by fixing the mass and allowing $q$ to vary, it
took only 250 generations to find the input model in 7 out of 10 runs.
This suggests that it might be more efficient to alternate between these
two subsets of 4 parameters, fixing the fifth parameter each time to its
best-fit value from the previous iteration, until the results of both fits
are identical. 

Since we do not know {\it a priori} the precise mass of the white dwarf,
we need to ensure that this iterative 4-parameter approach will work even
when the mass is initially fixed at an incorrect value. To test this, we
calculated the pulsation periods of the best-fit $0.65\ M_{\odot }$ model
for GD~358 from MNW [$T_{\rm eff}=22,600$~K, $\log(M_{\rm He}/M_*)=-2.74$,
20:80 C/O ``shallow''] and then iteratively applied the two 4-parameter
fitting routines, starting with the mass fixed at $0.60\ M_{\odot }$--- a
difference comparable to the discrepancy between the mass found by MNW and
the value found by \cite{bw94}. The series of fits leading to the input
model are shown in Table \ref{tab1}. This method required only 3
iterations, and for each iteration we performed 10 runs with different
random initialization to yield a probability of finding the best-fit much
greater than 99.9 percent. In the end, the method required a total of
$2.5\times 10^6$ model evaluations (128 trials per generation, 10 runs of
650 generations per iteration). This is about 200 times more efficient
than calculating the full grid in each iteration, and about 4,000 times
more efficient than a grid of the entire 5-dimensional space. 

Next, we applied this method to the observed pulsation periods of GD~358.
We initially fixed the mass at $0.61\ M_{\odot}$, the value inferred from
the original asteroseismological study by \cite{bw94}. The solution
converged after four iterations, and the best-fit values of the five
parameters were:
\begin{eqnarray*}
T_{\rm eff}           & = & 22,600\ {\rm K} \\
M_*/M_{\odot}         & = & 0.650\          \\ 
\log(M_{\rm He}/M_*)  & = & -2.74           \\
X_{\rm O}             & = & 0.84            \\ 
q                     & = & 0.49
\end{eqnarray*}
Note that the values of $M_*$, $T_{\rm eff}$ and $M_{\rm He}$ are
identical to those found in MNW. The best-fit mass and temperature still
differ significantly from the values inferred from spectroscopy by
\cite{bea99}. However, the luminosity of our best-fit model is consistent
with the luminosity derived from the measured parallax of GD~358
\citep{har85}. 

To alleviate any doubt that the GA had found the best combination of
$X_{\rm O}$ and $q$, and to obtain a more accurate estimate of the
uncertainties on these parameters, we calculated a grid of 10,000 models
with the mass, temperature, and helium layer mass fixed at their best-fit
values. A contour plot of this grid near the solution found by the GA is
shown in Figure \ref{fig1}.

\section{REVERSE\ APPROACH \label{REVSEC}}

The results of forward modeling with one adjustable point in the chemical
profile make it clear that information about the internal structure is
contained in the data, and we just need to learn how to extract it. If we
want to test more complicated adjustable profiles, forward modeling
quickly becomes too computationally expensive as the dimensionality of the
search space increases. We need to devise a more efficient approach to
explore the myriad possibilities. 

The natural frequency that dominates the determination of pulsation
periods in white dwarf models is the Brunt-V\"ais\"al\"a (BV) frequency,
which we calculate using the Modified Ledoux treatment described in
\cite{tfw90}. To 

\epsfxsize 3.5in
\epsffile{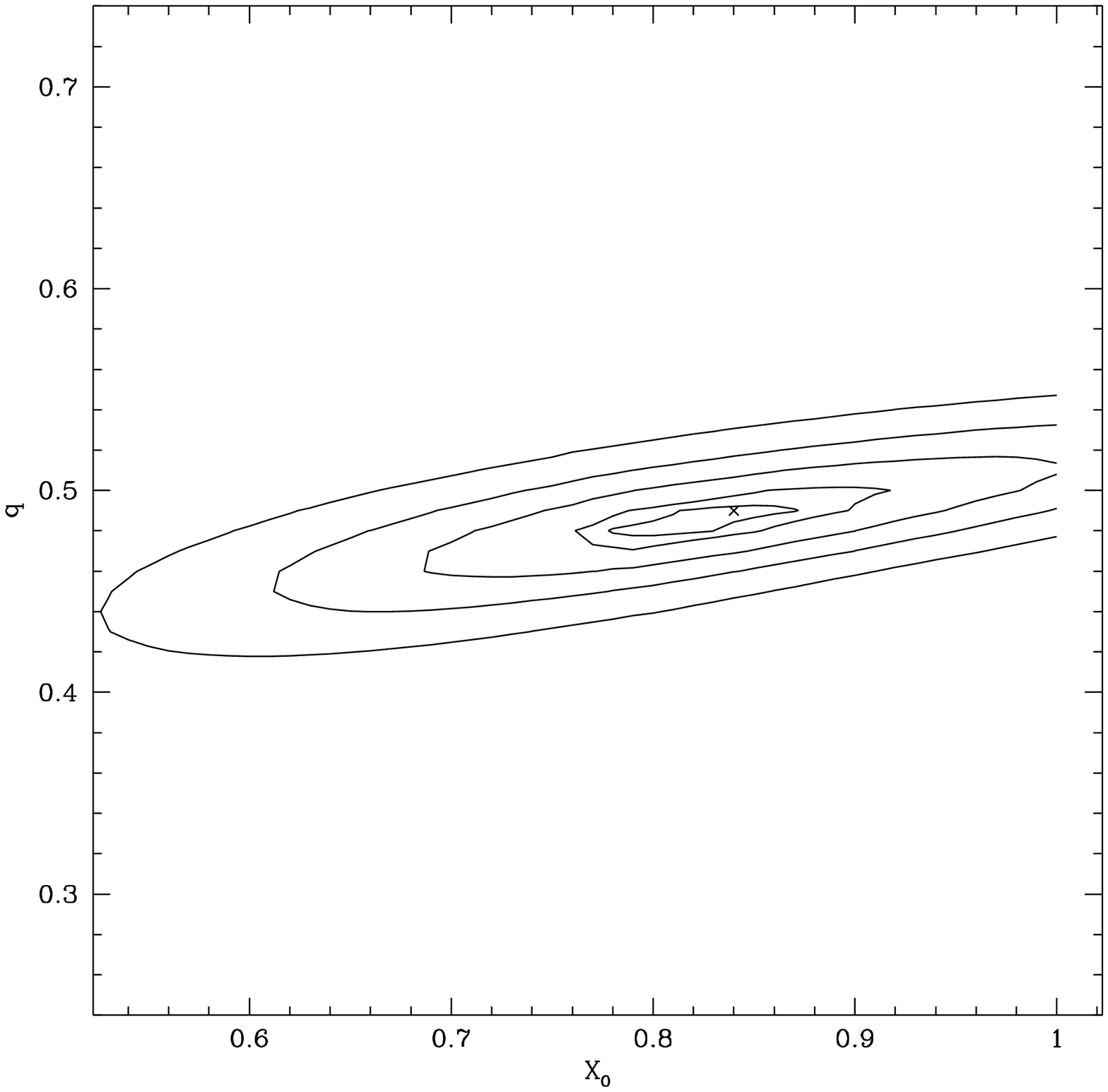}
\figcaption[f1.eps]{A contour plot of the central oxygen mass fraction
($X_{\rm O}$) versus the fractional mass location of the change in the
oxygen gradient ($q$), with $M_*$, $T_{\rm eff}$, and $M_{\rm He}/M_*$
fixed at their best-fit values. The model with the absolute minimum
residuals (identical to the best-fit found by the GA) is marked with an X.
The contours are drawn at 1, 3, 10, 25 and 40 times the observational
noise. \label{fig1}}
\vskip 12pt

\noindent get a sense of how the pulsation periods depend on the
BV frequency in various regions of the model interior, we added a smooth
perturbation to the best-fit model of GD~358 from MNW, moving it one shell
at a time from the center to the surface. The perturbation artificially 
decreased the BV frequency across seven shells, with a maximum amplitude 
of 10 percent. We monitored the effect on each of the pulsation periods as
the perturbation moved outward through the interior of the model. The results 
of this experiment for the pulsation periods corresponding to those observed 
in GD~358 are shown in Figure \ref{fig2}. 

Essentially, this experiment
demonstrates that the pulsation periods are sensitive to the conditions
all the way down to the inner few percent of the model. Since the
observational uncertainties on each period are typically only a few
hundredths of a second, even small changes to the BV frequency in the
model interior are significant.

The root-mean-square (rms) residuals between the observed pulsation
periods in GD~358 and those calculated for the best-fit from forward
modeling are still much larger than the observational noise. This suggests
that either we have left something important out of our model, or we have
neglected to optimize one or more of the parameters that could in
principle yield a closer match to the observations. To investigate the
latter possibility, we introduced {\it ad hoc} perturbations to the BV
frequency of the best-fit model to see if the match could be improved. 
Initially, we concentrated on the region of the Brunt-V\"ais\"al\"a curve
that corresponds to the internal chemical profile. 

If we look at the BV frequency for models with the same mass, temperature,
helium layer mass, and central oxygen mass fraction but different internal
chemical profiles (see Figure \ref{fig3}) it becomes clear that the
differences are localized. In general, we find that changes in the
composition gradient cause shifts in the BV frequency. Moving across an
interface where the gradient becomes steeper, the BV frequency shifts
higher; at an interface 

\epsfxsize 3.5in
\epsffile{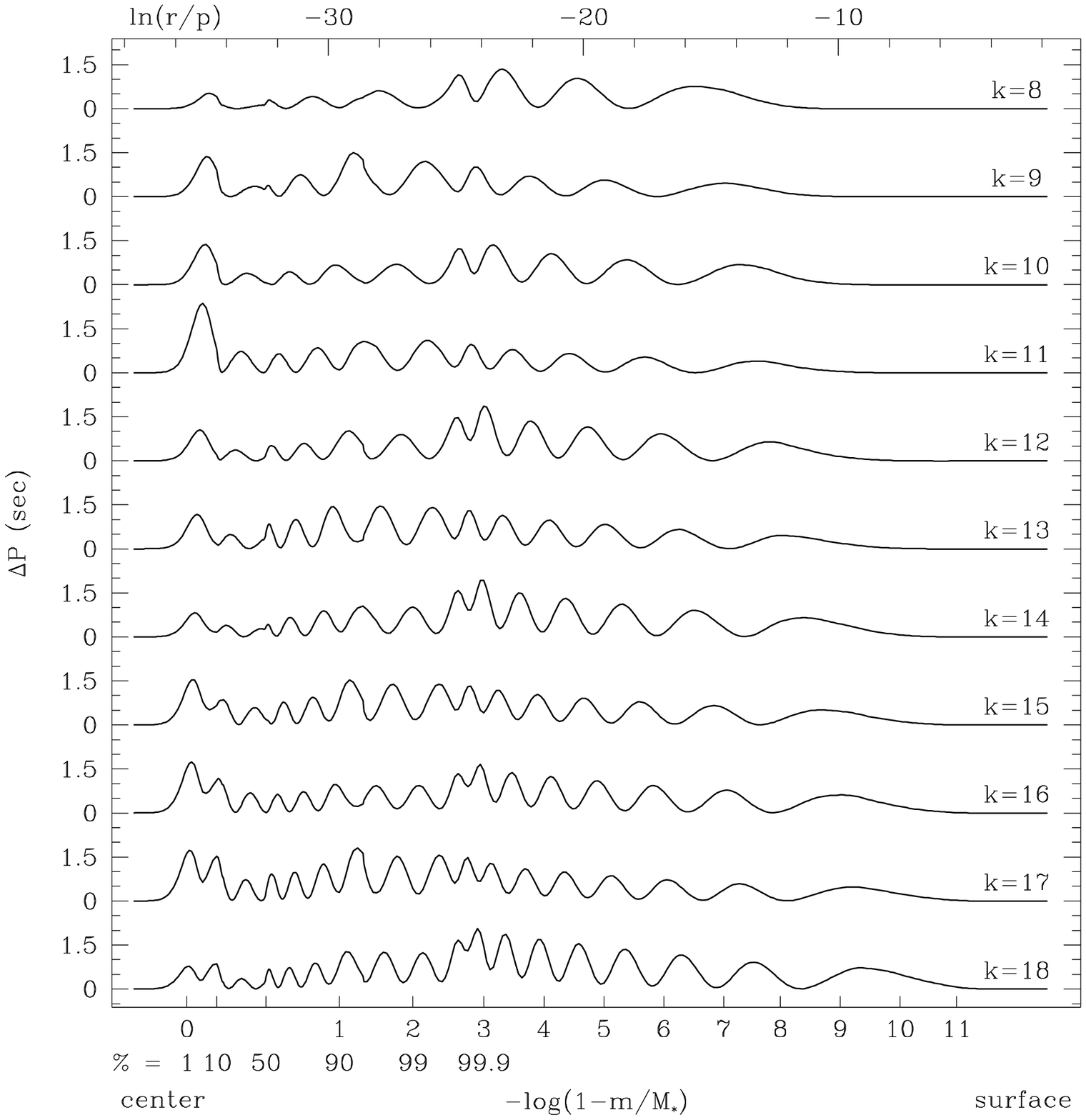}
\vskip 12pt
\figcaption[f2.eps]{For the best-fit model of GD~358 from \cite{mnw00}
this plot shows the change in pulsation period for $\ell=1$ modes of
various radial overtone number ($k$) which result from a smooth artificial 
10\% decrease in the Brunt-V\"ais\"al\"a frequency as a function of the natural
log of the ratio of the distance from the center of the model to the local
pressure (top axis) and the fractional mass coordinate $-\log(1-m/M_*)$
(bottom axis). The center of the model is to the left, and the surface is
to the right. Also indicated is the mass fraction expressed as a
percentage for several values closer to the center.  \label{fig2}}
\vskip 30pt

\epsfxsize 3.5in
\epsffile{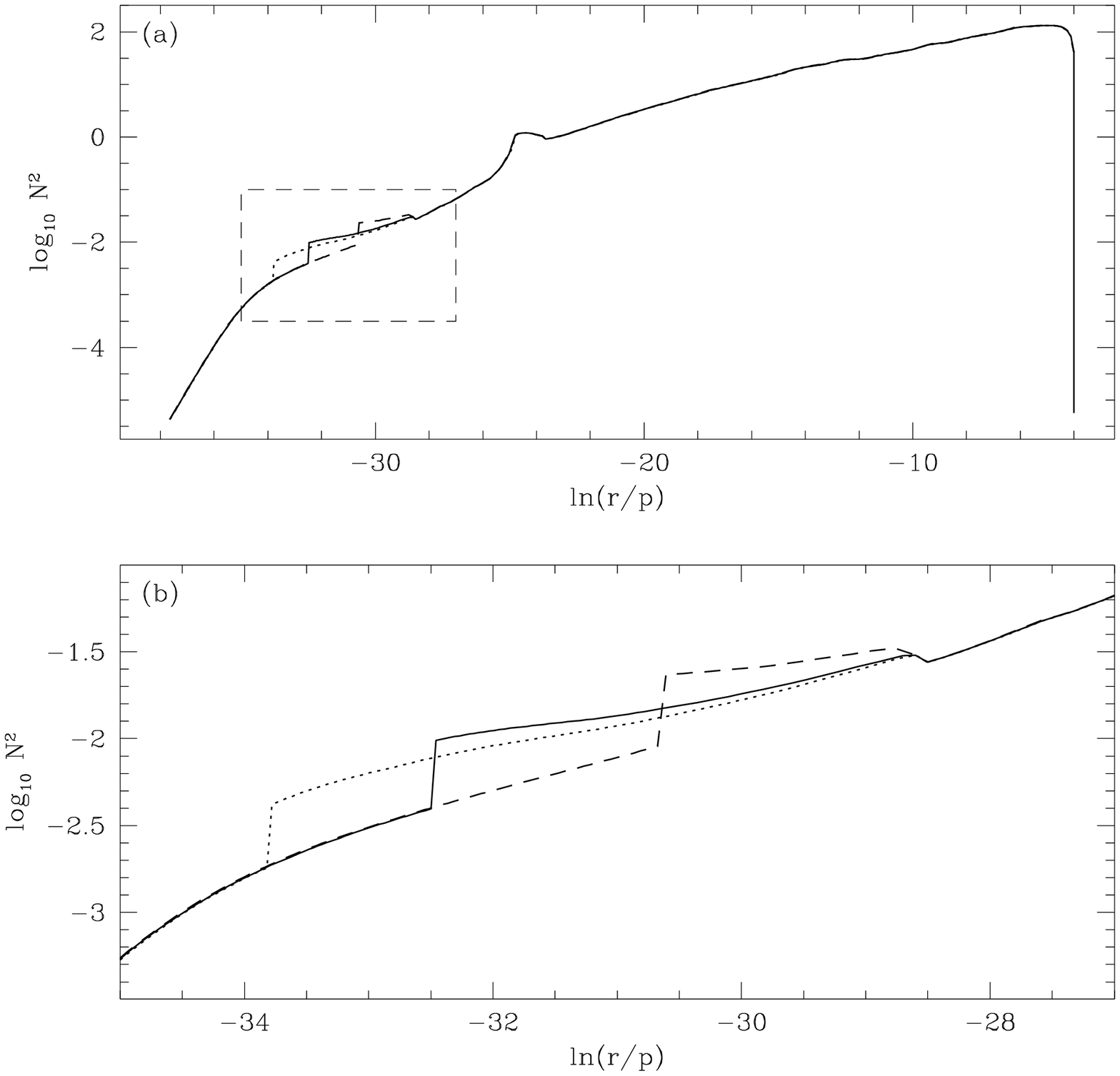}
\figcaption[f3.eps]{The Brunt-V\"ais\"al\"a frequency as a function of the
radial coordinate ln(r/p) for several models with the same mass,
temperature, helium layer mass, and central oxygen mass fraction but
different internal chemical profiles (a) from the center of the model at
left to the surface at right, and (b) only in the range of ln(r/p)
indicated by the dashed box in the upper panel. The three curves
correspond to a profile with $q$ equal to 0.2 (dotted), 0.49 (solid), and
0.8 (dashed).\label{fig3}}

\noindent where the gradient becomes more shallow, the BV
frequency shifts lower. The greater the change in the gradient, the larger
the shift in the BV frequency. 

\subsection{Proof of Concept}

We began by generating a model with the same mass, temperature, helium
layer mass, and internal composition as the best-fit from MNW, but using a
uniform internal chemical profile with constant 20:80 C/O out to the 95
percent mass point. We identified a sequence of 100 shells in this model
spanning a range of fractional mass from 0.20 to 0.97 and perturbed the BV
frequency to try to produce a better match to the observations. We
parameterized the perturbation as a multiplicative factor applied to the
BV frequency over a range of shells, described by four parameters: (1) the
innermost shell to perturb, (2) the magnitude of the perturbation at the
innermost shell, (3) the number of shells in the perturbation range, and
(4) the magnitude of the perturbation at the outermost shell.  These 4
parameters are sufficient to describe a profile with two abrupt changes in
the composition gradient. 

The innermost shell was allowed to be any of the 100 shells, and the
number of shells in the perturbation range was allowed to be between 0 and
100. If the chosen range would introduce perturbations outside of the
sequence of 100 shells, the outermost of these shells was the last to be
perturbed. The magnitude of the perturbation was allowed to be a
multiplicative factor between 1.0 and 3.0 at both the innermost and
outermost perturbed shells, and was interpolated linearly in the shells
between them. 

By using a GA to optimize the perturbation, we can try many random
possibilities and eventually find the global best-fit with far fewer model
evaluations than a full grid of the parameter space. We demonstrated this
by introducing a particular perturbation to the model and determining the
pulsation periods. Using the same unperturbed model, we then attempted to
find the parameters of the perturbation by matching the periods using the
GA. In 9 out of 10 runs (500 generations of 64 trials), the GA found a
first-order solution within two grid points of the input perturbation. 
Thus, the GA combined with a small (625 point) grid search yields a 90
percent probability of success for an individual run. By repeating the
process several times, the probability of finding the correct answer
quickly exceeds 99.9 percent, even while the number of model evaluations
required remains hundreds of times lower than a full grid of the
parameter-space. 

\subsection{Application to GD~358}

Having demonstrated that the method works on calculated model periods, we
applied it to the observed pulsation periods of GD~358. We began with a
model similar to the new best-fit determined from the forward modeling
described in \S \ref{FWDSEC}, but again using a uniform internal chemical
profile (constant 16:84 C/O out to 0.95 $m/M_*$). After the GA had found
the best-fit perturbation for GD~358, we reverse-engineered the
corresponding chemical profile.

To accomplish this, we first looked in the unperturbed model for the
fractional mass corresponding to the innermost and outermost shells in the
perturbation range. We fixed the oxygen abundance to that of the
unperturbed model from the center out to the fractional mass of the
innermost perturbed shell. The size of the shift in the BV frequency is
determined by how much the composition gradient changes at this point, so
we adjusted the oxygen abundance at the fractional mass of the outermost
perturbed shell until the change in the gradient produced the required
shift. Finally, we fixed the oxygen abundance to that value from the
outermost perturbed shell out to a fractional mass of 0.95, where it
abruptly goes to zero. 

After we found the C/O profile of the best-fit perturbation in this way,
we fixed this reverse-engineered profile in the models and performed a new
fit from forward modeling with the GA to re-optimize the mass,
temperature, helium layer mass, and central oxygen mass fraction. The BV
curve of the final model differs slightly, of course, from that of the
original uniform composition model with the perturbation added. But the
approximate internal structure is preserved, and leads to a better match
to the observed pulsation periods than we could have otherwise found.

\section{RESULTS}

The calculated periods and period spacings ($\Delta P = P_{k+1} - P_k$) 
for the best-fit models from the new forward modeling and from the reverse
approach are shown in the bottom two panels of Figure \ref{fig4} along
with the data for GD~358. The best-fit models of \cite{bw94} and MNW are
shown in the top two panels for comparison. The data in Figure \ref{fig4}
for the observations and our best-fit models are given in Table
\ref{tab2}. Some of the improvement evident in the panels of Figure
\ref{fig4} is certainly due to the fact that we have increased the number
of free parameters. To evaluate whether or not the new models represent a

\epsfxsize 3.5in
\epsffile{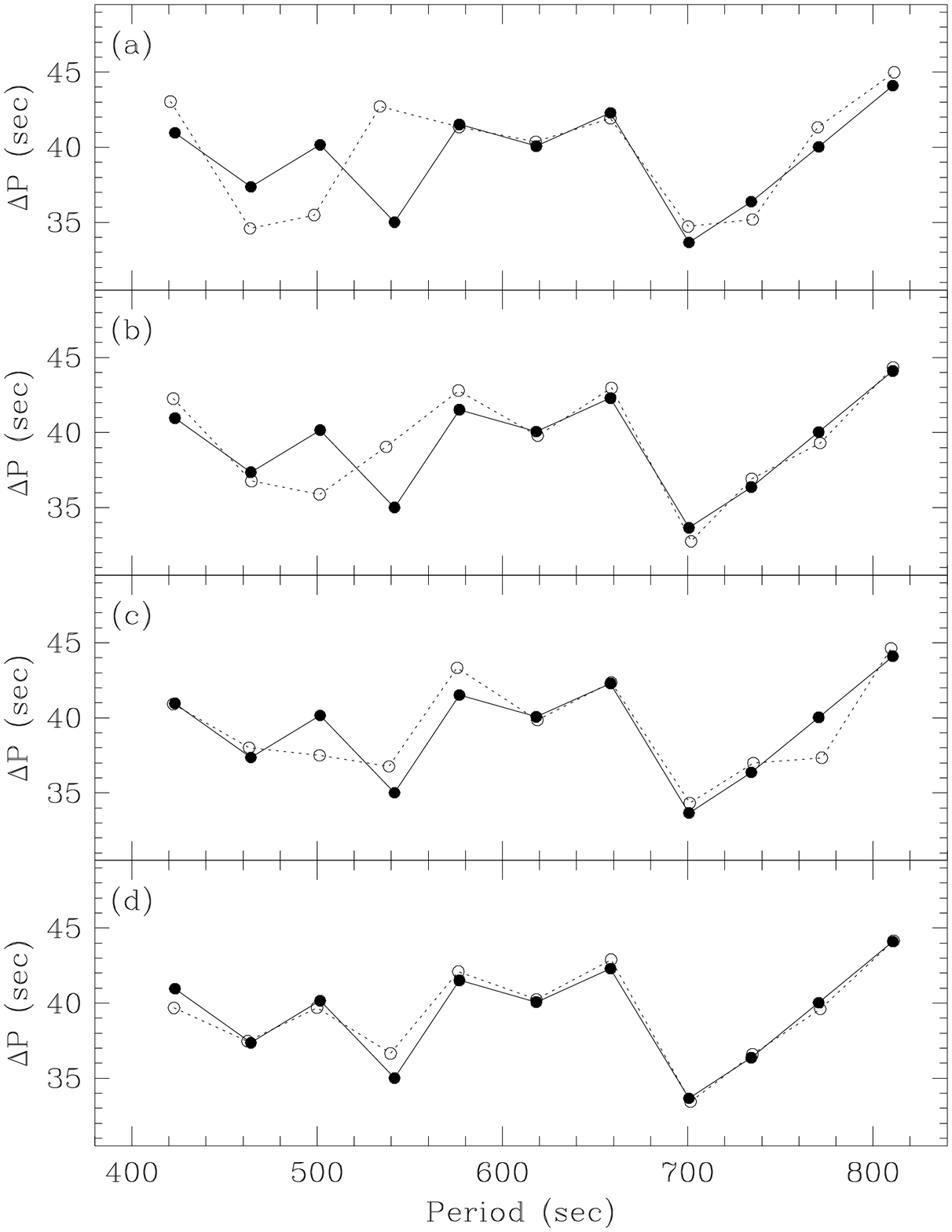}
\figcaption[f4.eps]{The periods and period spacings observed in GD~358
(solid points) with the theoretical best-fit models (open points)  from
(a) \cite{bw94}, (b) \cite{mnw00}, (c) the new forward modeling presented
in \S \ref{FWDSEC}, and (d) the reverse approach presented in \S
\ref{REVSEC}. Uncertainties on the observations are smaller than the size
of the points in this figure.\label{fig4}}

\begin{table*}
\begin{center}
\tablenum{1}\label{tab2}
\centerline{\sc Table 2}
\centerline{\sc Periods and Period Spacings for GD~358 and the Best-fit Models.}
\vskip 5pt
\begin{tabular}{lllccccccccc}
\hline\hline
&\multicolumn{2}{c}{Observed} &&\multicolumn{2}{c}{3-par fit} &
&\multicolumn{2}{c}{5-par fit}&&\multicolumn{2}{c}{7-par fit}     \\ 
\cline{2-3}\cline{5-6}\cline{8-9}\cline{11-12}
$k$ &$P$&$\Delta P$& &$P$&$\Delta P$& &$P$&$\Delta P$& &$P$&$\Delta P$  \\
\hline
8$\dotfill$& 423.27 & 40.96 && 422.31 & 42.26 && 422.36 & 40.92 && 422.75 & 39.69 \\
9$\dotfill$& 464.23 & 37.36 && 464.57 & 36.77 && 463.28 & 38.01 && 462.43 & 37.46 \\
10$\ldots$ & 501.59 & 40.16 && 501.35 & 35.88 && 501.29 & 37.50 && 499.90 & 39.70 \\
11$\ldots$ & 541.75 & 35.01 && 537.23 & 39.04 && 538.79 & 36.77 && 539.60 & 36.65 \\
12$\ldots$ & 576.76 & 41.52 && 576.27 & 42.79 && 575.56 & 43.33 && 576.25 & 42.10 \\
13$\ldots$ & 618.28 & 40.07 && 619.06 & 39.79 && 618.89 & 39.85 && 618.36 & 40.25 \\
14$\ldots$ & 658.35 & 42.29 && 658.85 & 42.97 && 658.74 & 42.36 && 658.61 & 42.90 \\
15$\ldots$ & 700.64 & 33.66 && 701.82 & 32.76 && 701.10 & 34.33 && 701.51 & 33.44 \\
16$\ldots$ & 734.30 & 36.37 && 734.58 & 36.92 && 735.42 & 36.99 && 734.95 & 36.59 \\
17$\ldots$ & 770.67 & 40.03 && 771.50 & 39.30 && 772.41 & 37.34 && 771.54 & 39.60 \\
18$\ldots$ & 810.7  & 44.1  && 810.80 & 44.34 && 809.75 & 44.63 && 811.14 & 44.15 \\
\hline\hline
\end{tabular}
\end{center}
\vskip -12pt
\end{table*}

\noindent {\it significant} improvement to the fit, we use the Bayes 
Information Criterion (BIC), following \cite{mmw01}. 

The fit in MNW used $n_p=3$ completely free parameters, though they also
sampled several combinations of two additional parameters. This amounts to
a partial optimization in 5 dimensions. To make a fair prediction of the
residuals required to be considered significant we use their best-fit
carbon core model, which represents the best truly 3-parameter fit. This
model had rms residuals of $\sigma(P)=2.30$ seconds for the periods and
$\sigma(\Delta P)=2.65$ seconds for the period spacings. For $N=11$ data
points, the BIC leads us to expect the residuals of a $n_p=5$ fit to
decrease to $\sigma(P)=1.84$ and $\sigma(\Delta P)=2.13$ just from the
addition of the extra parameters. In fact, the fit from the new forward
modeling presented in \S \ref{FWDSEC} has $\sigma(P)=1.28$ and
$\sigma(\Delta P)=1.42$, so we conclude that the improvement is
statistically significant. 

The results of the reverse approach presented in \S \ref{REVSEC} are
harder to evaluate because we are perturbing the BV frequency directly,
rather than through a specific parameter. We consider each additional
point in the internal chemical profile where the composition gradient
changes to be a free parameter. Under this definition, the perturbed
models are equivalent to a 7-parameter fit since there are three such
points in the profiles, compared to only one for the 5-parameter case. If
we again use the BIC, we expect the residuals to decrease from their
$n_p=5$ values to $\sigma(P)=1.03$ and $\sigma(\Delta P)=1.14$ seconds. 
After re-optimizing the other four parameters using the profile inferred
from the reverse approach, the residuals actually decreased to
$\sigma(P)=1.11$ and $\sigma(\Delta P)=0.71$ seconds. The decrease in the
period residuals is not significant, but the period spacings are improved
considerably. This is evident in the bottom panel of Figure \ref{fig4}. 

Adopting the central oxygen mass fraction from the best-fit forward
modeling, $X_{\rm O}=0.84\pm3$ we can place preliminary constraints on the
$^{12}$C$(\alpha ,\gamma )^{16}$O cross-section. \cite{sal97} made
detailed evolutionary calculations for main-sequence stellar models with
masses between 1 and 7 M$_{\odot}$ to provide internal chemical profiles
for the resulting white dwarfs. For the bulk of the calculations they
adopted the rate of \cite{cau85} for the $^{12}$C$(\alpha ,\gamma )^{16}$O
reaction ($S_{300}=240$ keV barns), but they also computed an evolutionary
sequence using the lower cross-section inferred by \cite{wtw93} from solar
abundances ($S_{300}=170\pm50$ keV barns). The chemical profiles from both
rates had the same general shape, but the oxygen abundances were uniformly
smaller for the lower rate. In both cases the C/O ratio was constant out
to the 50 percent mass point, a region easily probed by white dwarf
pulsations. 

The central oxygen mass fraction is lower in higher mass white dwarf
models. The rate of the triple-$\alpha$ reaction (a three-body process) 
increases faster at higher densities than does the $^{12}$C$(\alpha
,\gamma )^{16}$O reaction. As a consequence, more helium is used up in the
production of carbon, and relatively less is available to produce oxygen
in higher mass models. Interpolating between the models of \cite{sal97}
which used the higher value of the cross-section, we expect a central
oxygen mass fraction for a $M=0.65\ M_{\odot}$ model of $X_{\rm O}^{\rm
high}=0.75$. Using additional calculations for the low rate (M.~Salaris
2001, private communication), the expected value is $X_{\rm O}^{\rm
low}=0.62$. Extrapolating to the value inferred from our new forward
modeling, we estimate that the astrophysical S-factor at 300 keV for the
$^{12}$C$(\alpha , \gamma )^{16}$O cross-section is in the range
$S_{300}=290 \pm 15$ keV barns (internal uncertainty only). 

The internal chemical profiles corresponding to the best-fit models from
\S \ref{FWDSEC} and \S \ref{REVSEC} are shown in Figure \ref{fig5} with
the theoretical profile for a 0.61 $M_{\odot}$ model from \cite{sal97},
scaled to a central oxygen mass fraction of 0.80. The profile from the
best-fit forward modeling matches the location and slope of the initial
shallow decrease in the theoretical profile. The reverse approach also
finds significant structure in this region of the model, and is
qualitatively similar to the \cite{sal97} profile to the extent that our
parameterization allows.

\section{DISCUSSION \& FUTURE WORK}

The extension of the genetic-algorithm-based approach to optimize the
internal composition and structure of our best-fit white dwarf models to
GD~358 has yielded some exciting results. The values of the 3 parameters
considered in the initial study ($M_*, T_{\rm eff}, M_{\rm He}$) are
unchanged in the full 5-parameter fit, so we feel confident that they are
the most important for matching the gross period structure. The
significant improvement to the fit made possible by including $X_{\rm O}$
and $q$ as free parameters confirms that the observed pulsations really do
contain information about the hidden interiors of these stars. 

The efficiency of the GA relative to a grid search is much higher for the
larger parameter-space, and the ability of the method to find the global
best-fit is undiminished. The application of this method to data on
additional pulsating white dwarfs will help us to assess the significance
of our results, so we can begin to understand the statistical properties
of these ubiquitous and relatively simple stellar objects. It may also
provide us with

\epsfxsize 3.5in
\epsffile{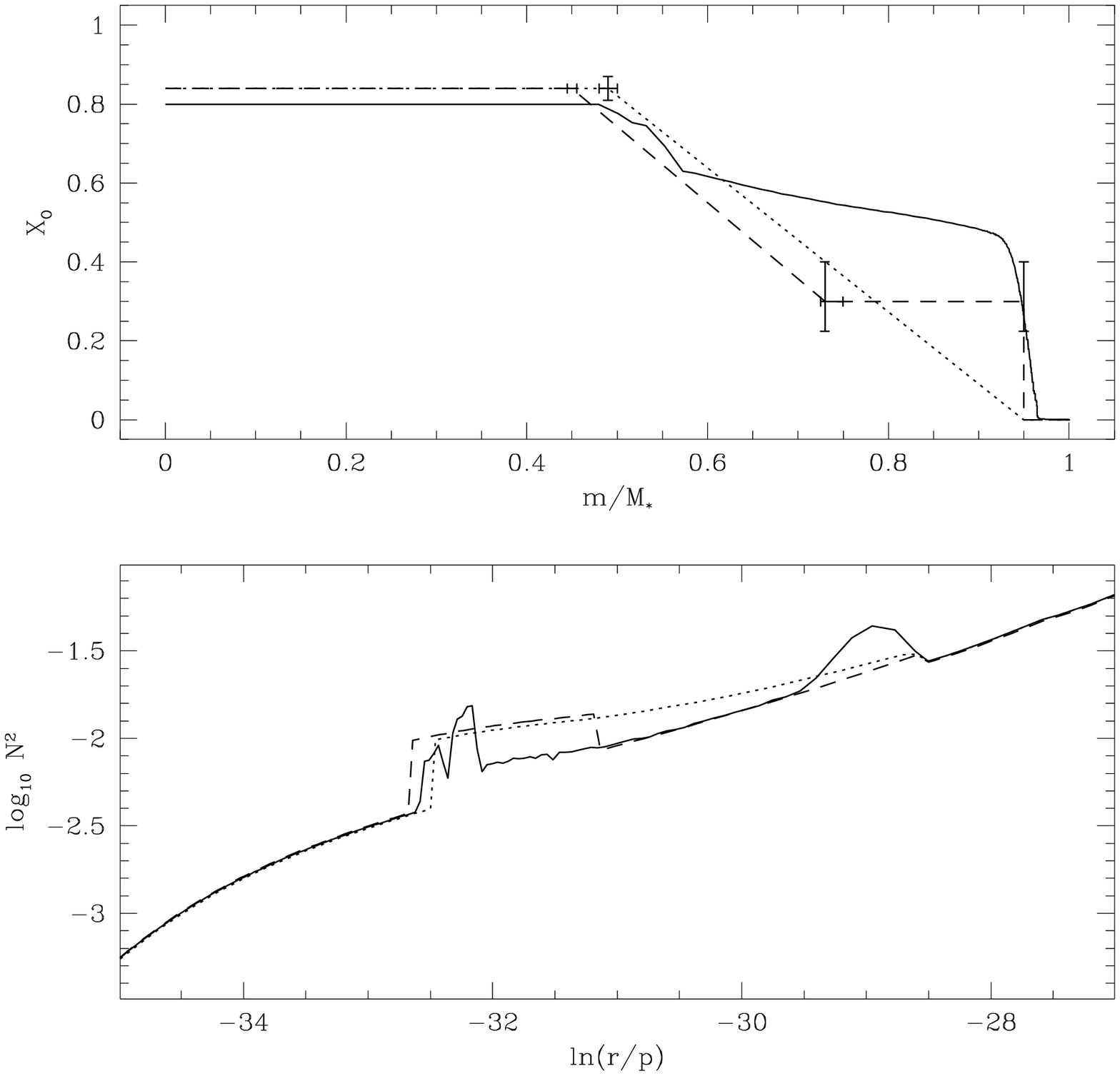}
\figcaption[f5.eps]{The internal oxygen profiles (top) and the
corresponding region of the Brunt-V\"ais\"al\"a curves (bottom) for the
best-fit forward model from \S \ref{FWDSEC} (dotted), the result of the
best-fit reverse approach from \S \ref{REVSEC} (dashed), and the scaled
theoretical calculations of \cite{sal97} for comparison (solid). 
\label{fig5}}
\vskip 12pt

\noindent new insights into the processes involved in white dwarf
evolution. 

Our reverse approach to model-fitting was originally motivated by the fact
that the variance of our best-fit model from the forward method was still
far larger than the observational uncertainties. Although this initial
application has demonstrated the clear potential of this approach to yield
better fits to the data, the improvement to the residuals was only
marginally significant. We will continue to develop this approach, but we
must simultaneously work to improve the input physics of our models. In
particular, since the internal profile from this initial application is
qualitatively similar to the theoretical profiles (given what is possible
with this parameterization), we should really include oxygen in our
envelopes and eventually calculate fully self-consistent models out to the
surface. 

Turning the central oxygen abundance into a more precise constraint on the
$^{12}$C$(\alpha ,\gamma )^{16}$O nuclear reaction cross-section will
require additional detailed simulations like those of \cite{sal97}.  By
determining the range of values for the cross-section that produce a
central oxygen abundance within the measurement uncertainties of $X_{\rm
O}$, we should be able to surpass the precision of the extrapolation from
laboratory data by nearly an order of magnitude. To assess the systematic
uncertainties, we will also need to determine the relative importance of
model-dependent details such as convection.

\acknowledgements 

We would like to thank Ed Nather and Mike Montgomery for helpful
discussions, and M.~Salaris for providing us with data files of the
theoretical white dwarf internal chemical profiles. We are grateful to the
High Altitude Observatory Visiting Scientist Program for fostering this
collaboration in a very productive environment for two months during the
summer of 2000. This work was supported by grant NAG5-9321 from the
Applied Information Systems Research Program of the National Aeronautics
\& Space Administration, and in part by grant AST-9876730 from the
National Science Foundation. 

\epsfxsize 3.5in
\epsffile{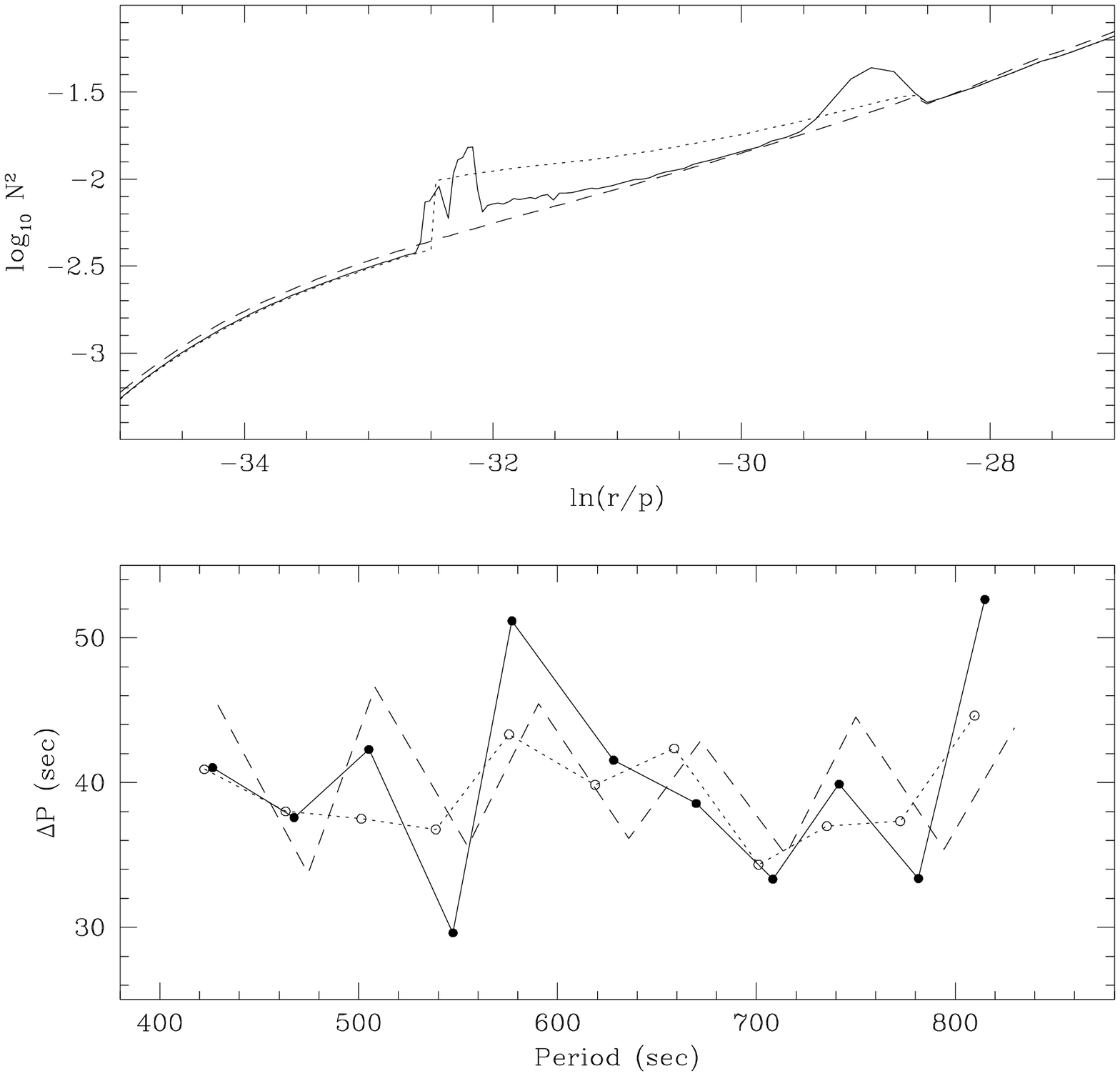}
\figcaption[f6.eps]{The Brunt-V\"ais\"al\"a frequency in the region
of the models that correspond to the internal chemical profile (top panel)
and the period spacing diagrams (bottom panel) for our best-fit forward 
model from \S 2 (dotted), a uniform 20:80 C/O model (dashed), and a
scaled theoretical profile from \cite{sal97} for comparison (solid).
\label{fig6}}

\appendix

At first glance, it may seem that our parameterization of the internal 
chemical profile might introduce large deviations from the mean period 
spacing due to the abrupt change in the composition gradient. To
demonstrate that the models are not unusually sensitive to our profiles
when compared to smooth profiles, we calculated the pulsation periods
of models with the best-fit parameters from \cite{mnw00} using several
internal chemical profiles. The results are shown in Figure \ref{fig6}. 
In the top panel, we show the Brunt-V\"ais\"al\"a frequency in the region
of the models that correspond to the internal chemical profile. In the
bottom panel, we show the period spacing diagrams for each of these
models. We find that deviations from the mean period spacing for a
model using an internal chemical profile from our parameterization have
a magnitude comparable to those caused by a profile with a uniform C/O
mixture out the the 95 percent mass point. The smooth profile from
\cite{sal97} causes the largest deviations of the three.

\end{document}